\documentclass{ws-procs10x7_preprint}
\usepackage{balance}

\newcolumntype{d}[1]{D{.}{.}{#1}}

\def\MET{\mbox{${\hbox{$E$\kern-0.6em\lower-.1ex\hbox{/}}}_T$ }} 

\makeindex
\begin{document}

\title{ $t\overline{t}$ Production at the Tevatron: Event Selection and Cross Section Measurement}

\author{D.C. O'Neil$^*$ \\ (for the CDF and D\O\ collaborations)}

\address{Department of Physics, Simon Fraser University, Burnaby, B.C. V5A 1S6, Canada\\$^*$E-mail: doneil@sfu.ca}


\twocolumn[\maketitle\abstract{The Fermilab Tevatron is currently the only
    collider capable of producing and studying top quarks. The dominant
    mechanism for top quark production at the Tevatron is $t\overline{t}$
    production via the strong interaction. The precise measurement of the
    cross section of this process is a test of the QCD prediction. In Run II
    of the Tevatron it should be possible to achieve an experimental error on
    this cross section which is comparable or better than the current
    theoretical precision. This paper presents the basic event selection
    criteria for $t\overline{t}$ events at the Tevatron and the latest
    measurements of the $t\overline{t}$ cross section. }

 ]

\section{Introduction}

The Fermilab Tevatron is currently producing 1.96~TeV centre-of-mass
$p\overline{p}$ collisions at unparalleled luminosities for the CDF and D\O\
experiments. This data-taking period, known as Run II, began in 2001. The
datasets obtained by each experiment now exceed an integrated luminosity of
1fb$^{-1}$, approximately a factor of 10 more than the Run I dataset. This new
dataset allows precision measurements of the properties of the top
quark. In particular, the cross section for top quark production is now
measured to unprecedented precision, allowing tests of the QCD prediction and
searches for new physics. 

At the Tevatron, the dominant mechanism for top quark production is pair
production via the strong interaction. The theoretical cross section for top
pair production at the Tevatron is calculated at NLO as $6.8$pb\cite{NLOcross}
with a precision of 10-12\%. The standard model also predicts the production of
single top quarks via the electroweak interaction, however, this process has
not yet been observed and is not addressed in this paper.

In addition to a general description of top quark event selection in each
analysis channel, a selected cross section measurement for each channel in each
experiment at the time of ICHEP 2006 is presented here. All of these
results have been presented at conferences previously except for the CDF
all-hadronic result which is new for ICHEP 2006.  

\section{Top Quark Decays}
Among quarks, the top quark is unique in that it decays before it can
hadronize. That decay is to a $W$-boson and a $b$-quark virtually 100\% of the
time. The decay of the 2 $W$-bosons produced in $t\overline{t}$ decay then
defines the different analysis channels pursued at the Tevatron: dilepton, in
which both $W$s decay to an electron or muon; lepton + jets, in which one $W$
decays to an electron or muon and the other decays hadronically; and
all-hadronic, in which both $W$s decay hadronically. The relative branching
fractions of these channels are approximately 4\%, 30\%, 45\%
respectively. Separate analyses and cross section measurements are performed
in each of these channels by both D\O\ and CDF.
%

\section{Selection of Top Quark Events}

While it is necessary to tailor selection criteria for each $t\overline{t}$
channel, there are certain general features of top quark events which can be
exploited in all channels. 

Thresholds on the number and transverse momentum of leptons and jets in each
event are used in all selection schemes. Top quark events generally have a
high multiplicity of high-$p_T$ jets. The scalar sum of the $p_T$ of all of
these objects (including missing transverse energy), known as $H_T$ is also a
commonly used variable in these analyses as top quark events tend to have
higher $H_T$ than background events. 

Some analyses presented in this paper also make use of b-jet
tagging. $t\overline{t}$ events contain at least 2 jets arising from
b-quarks. These jets can be identified (with some inefficiency) using
secondary vertex tagging since the B mesons travel some distance in the
tracking detectors of CDF and D\O\ before decaying.  

\section{Dilepton Channel}
The dilepton channel includes three final states: $ee$, $e\mu$ and
$\mu\mu$. The advantage of this channel is the clean signature consisting of 2
high-$p_T$ leptons, large missing transverse energy (\MET) and 2 b-jets. The
principle disadvantages are the low branching ratio and the presence of 2
neutrinos in each event. The principle background to dilepton events is $Z$
decay to 2 leptons with smaller background contributions from di-boson,
$W$+jet and QCD multijet production.  

\subsection{Event Selection and Cross Section Measurement}

\begin{figure}
\includegraphics[width=0.45\textwidth]{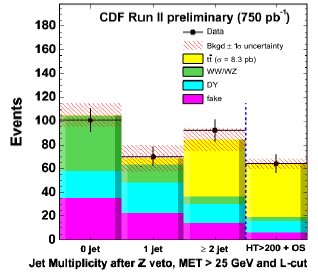}
\caption{CDF yields and background composition in dilepton channel as a function of jet multiplicity. The
  last bin includes the $H_T$ threshold and opposite-sign lepton requirement. \label{fig-cdfdilep}}
\end{figure}

The CDF cross section measurement in the dilepton channel is performed on
750pb$^{-1}$ of data. The selection requires 2
oppositely charged leptons with $E_T>20$GeV (one with ``tight'' selection, one loose), at least 2 jets with
$E_T>15$GeV, \MET $>$25~GeV (or $>$50~GeV is any lepton or jet is closer than
20$^\circ$ from the \MET direction), high \MET significance for events in the
Z mass region, and
$H_T>200$GeV. The yields after this selection are shown in
Figure~\ref{fig-cdfdilep}. The cross
section is extracted using:
\begin{equation}
\label{eq:cdfcross}
\sigma_{t\overline{t}} = \frac{N_{obs}-N_{bkg}}{\sum_i A_i {\cal L}_i}
\end{equation}
where $A$ is acceptance and ${\cal L}$ is luminosity, and is found to be:
\begin{displaymath}
 \sigma_{t\overline{t}}=8.3\pm 1.5 (stat) \pm 1.0 (syst) \pm 0.5 (lumi)pb.
\end{displaymath}

\begin{figure}
\includegraphics[width=0.45\textwidth]{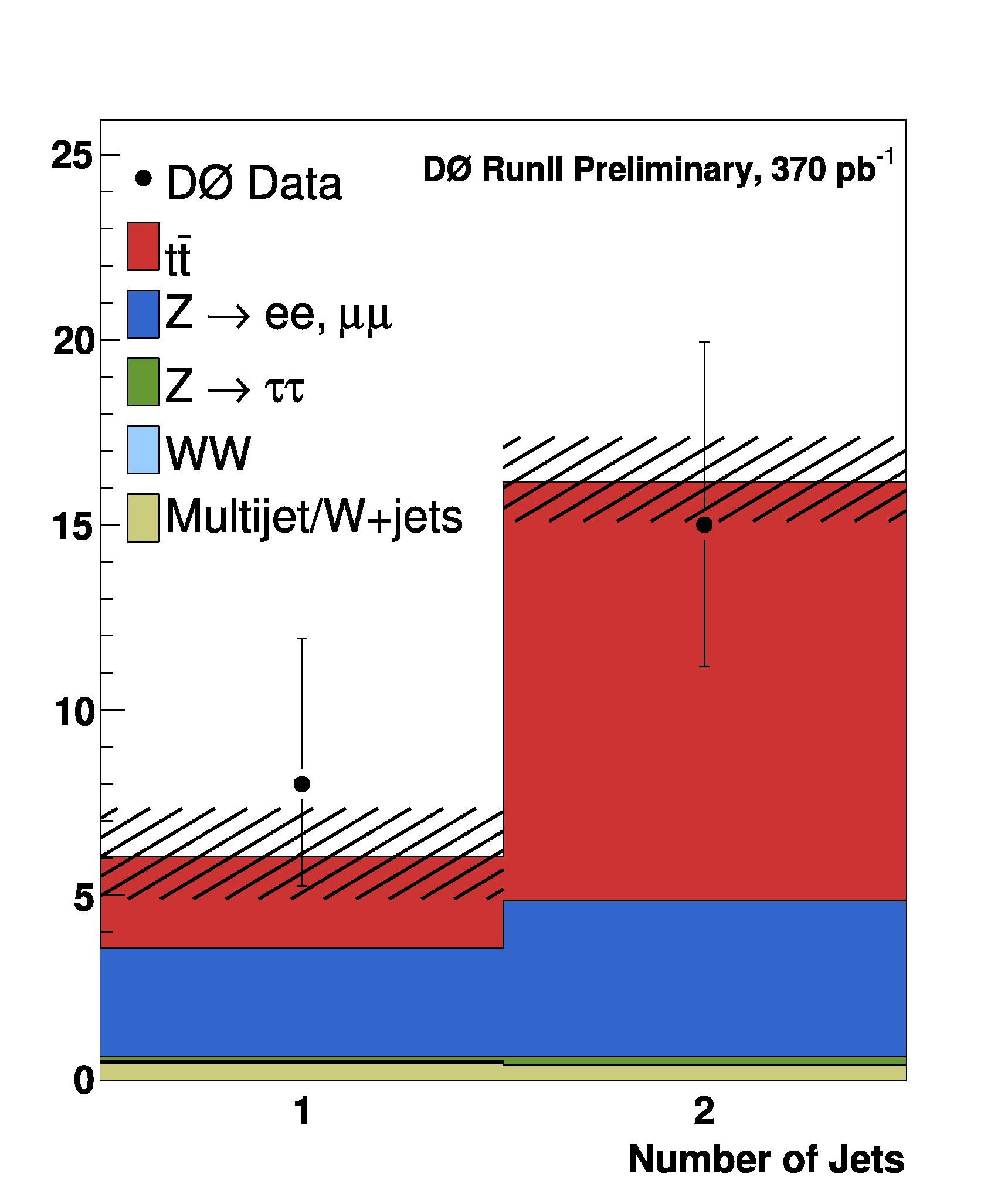}
\caption{D\O\ yields and background composition in dilepton channel as a function of jet multiplicity  \label{fig-d0dilep}}
\end{figure}

The D\O\ cross section measurement in the dilepton channel is performed on
370pb$^{-1}$pb of data. The selection requires 1 high-$p_T$ lepton and 1
isolated track of opposite charge, each with $E_T>15$GeV. Requiring only an
isolated track, rather than an object passing full lepton-ID, for the second
lepton in each event increases the signal acceptance. The analysis also
requires at least 1 jet with $E_T>20$GeV, \MET thresholds of at least 15~GeV
(more stringent requirements are made in the Z mass region and different
requirements are made for electron and muon channels),
at least one b-tag and an explicit veto of the $e\mu$ final state (which is
then combined for the final measurement). Figure~\ref{fig-d0dilep} shows the
yields after this selection. This sample leads
to a measured cross section of:
\begin{displaymath}
\sigma_{t\overline{t}}=8.6^{+1.9}_{-1.7}(stat) \pm 1.1 (syst) \pm 0.6 (lumi)pb. 
\end{displaymath}

\section{Lepton + Jets Channel}

The signature for the lepton + jets channel is exactly 1 high-$p_T$ lepton, 2
high-$p_T$ b-jets, 2 high-$p_T$ light-quark jets and missing transverse
energy. This channel generally provides the most stringent constraints on the
cross section as it is a compromise between branching ratio and clean signal. The leading background for this channel is $W$+jets production with
lesser contributions from QCD multijet, di-boson and $Z$ events.  

\subsection{Event Selection and Cross Section Measurement}

\begin{figure}
\includegraphics[width=0.45\textwidth]{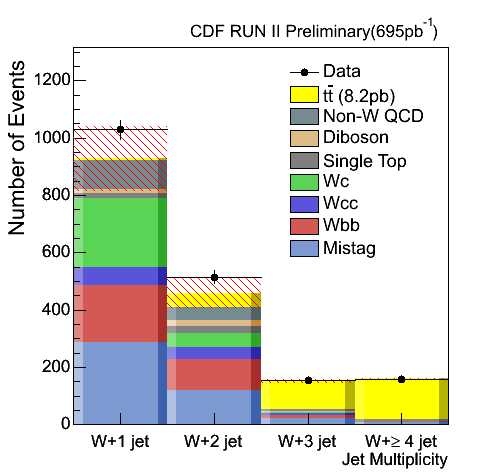}
\caption{CDF yields in lepton+jets channel as a function of jet multiplicity  \label{fig-cdflepjets}}
\end{figure}

The CDF analysis in the lepton+jets channel is performed on 695pb$^{-1}$ of
data. It requires 1 isolated lepton with $E_T>20$GeV, at least 3 jets
with $E_T>15$GeV, \MET$> 20$GeV, exactly 1 b-tagged jet and
$H_T>200$GeV. Figure~\ref{fig-cdflepjets} shows the event yield after
selection as a function of jet multiplicity. The 3 and 4 jet bins are
dominated by signal. The cross section is then extracted using equation~\ref{eq:cdfcross}
giving:
\begin{displaymath}
\sigma_{t\overline{t}}= 8.2 \pm 0.6 (stat) \pm 1.0 (syst)pb
\end{displaymath}
 This is the world's best single
measurement of the $t\overline{t}$ cross section. Events with 2 b-tagged jets
are not included in this result and are instead used as a cross-check
sample. The result from the cross-check sample agrees with that of the single-tag
sample.

\begin{figure}
\includegraphics[width=0.45\textwidth]{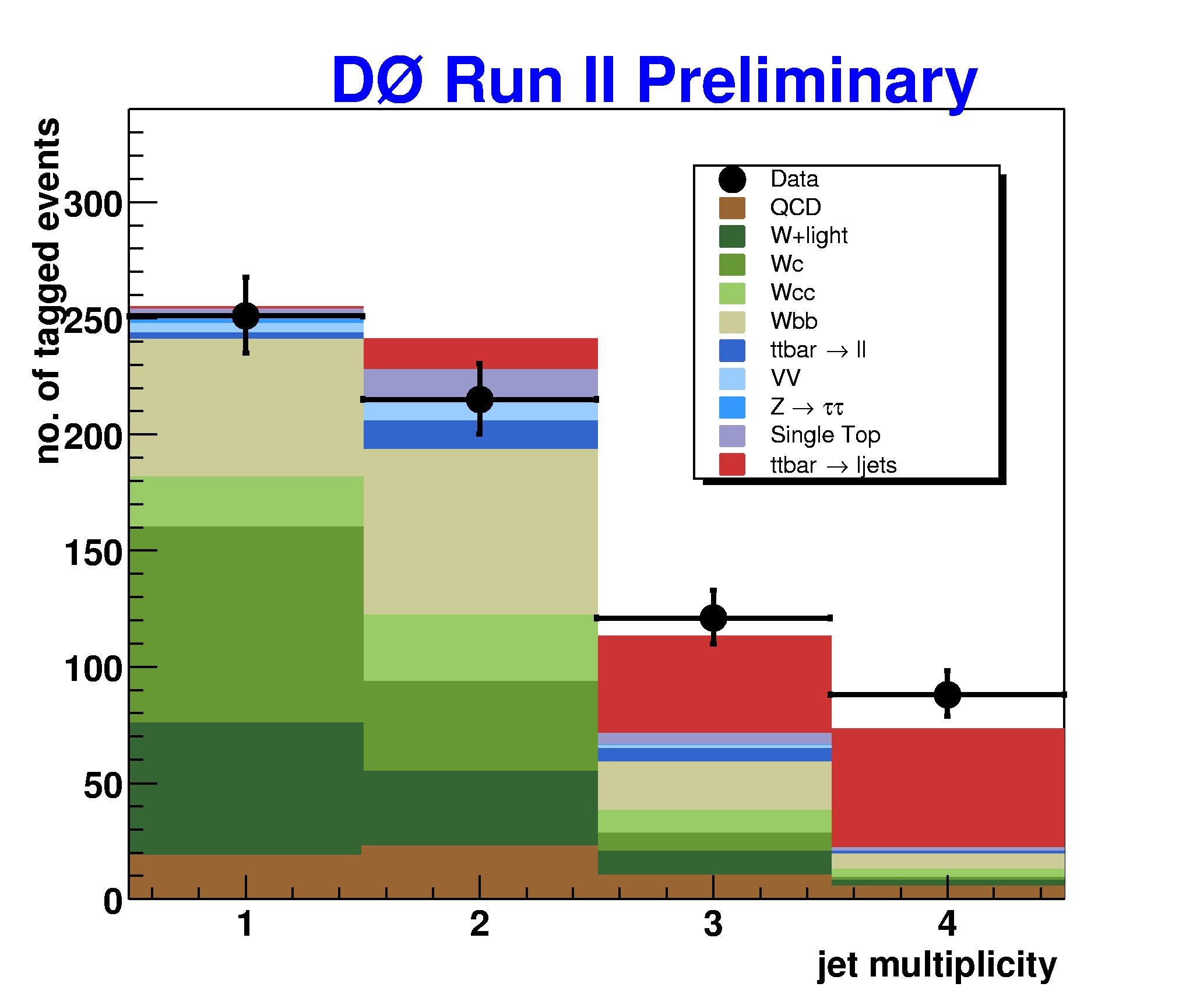}
\caption{D\O\ yields in lepton+jets channel as a function of jet multiplicity  \label{fig-d0lepjets}}
\end{figure}

The D\O\ analysis is performed on 370pb$^{-1}$ of data. It requires exactly 1 isolated central lepton with $E_T>20$GeV,
at least 1 jet with $E_T>15$GeV, \MET$>20$GeV and either 1 or 2 b-tagged
jets. Figure~\ref{fig-d0lepjets} shows the event yield after selection and
again it is clear that the 3rd and 4th jet bins are dominated by
signal. The cross section is extracted by fitting a liklihood function:
\begin{equation}
{\cal L} = \prod_i P(N_i^{obs},N_i^{pred}(\sigma_{t\overline{t}}))
\end{equation} 
where this is a product of probabilities to observe $N_i^{obs}$ given a
predicted number of events ($N_i^{pred}$) which is sensitive to the
$t\overline{t}$ cross section. This yields a cross section result:
\begin{displaymath}
\sigma_{t\overline{t}}= 8.1^{+1.3}_{-1.2} (stat+syst) 0.5 (lumi)pb.
\end{displaymath}
 This is the best current
$t\overline{t}$ cross section measurement from D\O\ . 

\section{All-Hadronic Channel}

The signature for the all-hadronic channel is 6 high-$p_T$ jets, 2 of them
from b's. The background is primarily from QCD multijet events with a smaller
contribution from $W$+jets. The advantages of this channel are the high
branching fraction and the lack of any neutrinos in the final state. The
disadvantage of this channel is the dependence on jet energy scale and the
high background rate.

\subsection{Event Selection and Cross Section Measurement}

\begin{figure}
\includegraphics[width=0.45\textwidth]{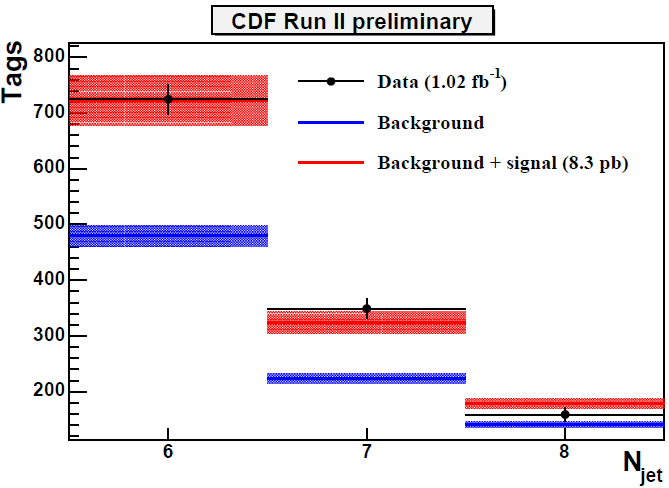}
\caption{CDF yields in All-hadronic channel as a function of jet multiplicity  \label{fig-cdfallhad}}
\end{figure}

The CDF all-hadronic analysis is performed on 1.02fb$^{-1}$ of data and was
first shown publicly at ICHEP 2006. It requires a veto on isolated leptons and
large \MET and 6-8 jets with $E_T>15$GeV. A topological neural network is then
constructed based on 11 inputs and 1 or more b-jets is
required. Figure~\ref{fig-cdfallhad} shows the yield as a function of the jet
multiplicity and demonstrates the significant signal present in the 6-8 jet
bins. The cross section is extracted based on the number of tags in the
data, rather than the number of events and yields: 
\begin{displaymath}
\sigma_{t\overline{t}}=8.3 \pm 1.0(stat)^{+2.3}_{-1.9}pb \pm 1.5 (lumi)pb.
\end{displaymath}

The most recent D\O\ measurement in the all-hadronic channel would be better
classified as a ``multi-jet'' analysis. This terminology is used since there is
no veto on isolated leptons or \MET in the analysis. Therefore, significant
signal from, for example, $t\rightarrow \tau\nu$ is present in this
sample. The analysis is performed on 360pb$^{-1}$ of data. It requires 6 jets,
with varying $p_T$ requirement, all of which must have at least 2 tracks
pointing to the event primary vertex and exactly 2 of which must be
b-tagged. The cross section extracted from this sample is:
\begin{displaymath}
\sigma_{t\overline{t}}= 12.1 \pm 4.9(stat) \pm 4.6(syst)pb.  
\end{displaymath}

\section{Summary}
Event selection techniques and cross section results for $t\overline{t}$
production at the Tevatron are presented for up to 1fb$^{-1}$ of integrated
luminosity. The results from several independent channels for both D\O\ and
CDF are presented and are consistent both with each other and with theory. The
best precision cross section measurements now rival theoretical precision.

\end{document}